\documentclass[aps,prd,showpacs,twocolumn]{revtex4-1}
\usepackage{amssymb}
\usepackage{amsmath}
\usepackage{amsfonts}
\usepackage{float}
\usepackage{accents}
\usepackage{graphicx}
\usepackage{epstopdf}
\usepackage{ulem}
\usepackage[colorlinks=true,pdfstartview=FitV,linkcolor=blue,citecolor=blue,urlcolor=blue,breaklinks=true]{hyperref}
\usepackage{color}
\usepackage[usenames,dvipsnames]{xcolor}
\usepackage{todonotes}

\begin{document}

\title{Lorentz-violating nonminimal coupling contributions in mesonic
hydrogen atoms and generation of photon higher-order derivative terms}
\author{R. Casana}
\email{rodolfo.casana@gmail.com}
\author{J. S. Rodrigues}
\email{josberg.silva@ufma.br}
\affiliation{Departamento de F\'{\i}sica, Universidade Federal do Maranh\~{a}o, Campus
Universit\'{a}rio do Bacanga, 65080-805, S\~{a}o Lu\'{\i}s, Maranh\~{a}o,
Brazil}
\author{F. E. P. dos Santos}
\email{frederico.santos@ufma.br}
\affiliation{Coordena\c{c}\~{a}o do Curso Interdisciplinar em Ci\^{e}ncia e Tecnologia,
Universidade Federal do Maranh\~{a}o, Campus Universit\'{a}rio do Bacanga,
65080-805, S\~{a}o Lu\'{\i}s, Maranh\~{a}o, Brazil.}

\begin{abstract}
We have studied the contributions of Lorentz-violating CPT-odd and CPT-even nonminimal couplings to the energy spectrum of the mesonic hydrogen and the higher-order radiative corrections to the effective action of the photon sector of a Lorentz-violating version of the scalar electrodynamics. By considering the complex scalar field describes charged mesons (pion or kaon), the non-relativistic limit of the model allows to attain upper-bounds by analyzing its contribution to the mesonic hydrogen energy. By using the experimental data for the $1S$ strong correction shift and the pure QED transitions $4P \rightarrow 3P$, the best upper-bound for the CPT-odd coupling is $<10^{-12}\text{eV}^{-1}$ and for the CPT-even one is $<10^{-16}\text{eV}^{-2}$. Besides, the CPT-odd radiative correction to the photon action is a dimension-5 operator which looks like a higher-order Carroll-Field-Jackiw term. The CPT-even radiative contribution to the photon effective action is a dimension-6 operator which would be a higher-order
derivative version of the minimal CPT-even term of the standard model extension. 
\end{abstract}

\pacs{11.30.Cp, 11.10.Gh, 11.15.Tk, 11.30.Er}
\maketitle

\section{Introduction}

The possibility of Lorentz and CPT violation in quantum electrodynamics (QED) has been studied intensely in the last years. The principal motivation is the possibility of occurring the spontaneous breaking of both symmetries at very high energy, i.e., the Planck scale \cite{Samuel}. The standard model extension (SME) \cite{Colladay} is the main framework proposed to study the possible effects of Lorentz violation into the standard model of the fundamental particles and their interactions. In the minimal SME the Lorentz-violating (LV) coefficients concerning to photonic and fermionic sectors are added in such a way to preserve gauge symmetry and renormalizability and, some of them have strong experimental upper bounds \cite{Data}. Detailed studies about the LV coefficients of the fermionic
sector can be founded in Refs. \cite{fermion}; {for the CPT-odd photon sector in Refs.\cite{Adam,Cherenkov1,photons1} while for the CPT-even  one in Refs. \cite{KM1,Risse,Cherenkov2}}. The introduction of Lorentz violation by means of operators with {mass dimension} higher than 4 was first considered in Ref. \cite{Myers} by studying the effects {of  a dimension-five operator on the particle dispersion relation.} Others applications of such a proposal were performed in Refs. \cite{Myers2}. The nonminimal SME embraces higher-order operators which were
considered, for example, in both the photon and fermion sectors in  Refs.
\cite{Kostelec,KMh}. Some implications  of the nonminimal SME  are
studied in Refs. \cite{Cambiaso}.  Another way to introduce Lorentz
violation in systems with scalar and/or fermion and gauge fields can be made
via nonminimal couplings which could  introduce or generate
higher-order  operators terms \cite{NM1,BaetaR2,Frede1, Frede2, Jonas1}.
Some consequences or  effects of those nonrenormalizable
coefficients have been investigated in several distinct scenarios \cite%
{NM3,NMhall, NMmaluf,NMbakke,NMABC,Radio1,Radio2,Radio21,Gomes, Aether,
Tiago2, BaetaR1,BaetaR3, Kostelecky2002,RadioFred}.

Some aspects of bound-states involving spinless particles can be treated
through the Klein-Gordon (KG) equation, while the quantum field theory
describing the interaction between spinless charged particles and the
electromagnetic field is known as scalar quantum electrodynamics (sQED). The
Lorentz-violating extensions of sQED and Klein-Gordon equation has
not received same attention like its fermionic version. The reason could be
the fact does not exist in standard model a fundamental spinless charged
particle or because it is not so rich phenomenologically like its fermionic
version. As an example, a CPT-odd power-counting renormalizable term is only
possible in photonic and fermionic sectors. Even so, some Lorentz-violating
{proposals for} the scalar electrodynamics has been studied in recent
years. For example, aspects like causality, unitarity and spontaneous
symmetry breaking of a LV and CPT-odd sQED were analyzed in Ref. \cite{Baeta}%
. The Higgs mechanism in the context of a Lorentz-violating and CPT-even
sQED was studied in Ref. \cite{ALTSCHUL}. At classical level many studies
about existence of vortices BPS in Lorentz-violating scalar electrodynamics
were performed in Refs. \cite{VORTEX}.

The aim of the manuscript is to analyze the contributions of
Lorentz-violating nonminimal couplings (CPT-odd and CPT-even ones) to the
energy-bound states of the mesonic hydrogen and to the effective action of
the photonic sector of the sQED. The LV nonminimal couplings are introduced
by modifying of the usual covariant derivative coupling the charged scalar
field and the Abelian gauge field. The manuscript is presented in the
following way: In Sec. II, we propose the model used for develop our
proposal. In Sec. III, our first goal is attained by analyzing the
nonrelativistic contribution of the nonminimal couplings to the Hamiltonian
of the mesonic hydrogen. The use of the experimental data from the $1S$
strong correction shift of the ground state and the $4P\rightarrow 3P$ pure
QED transitions of the mesonic hydrogen have allowed to impose upper-bounds
to the both CPT-odd and CPT-even coefficients. In Sec. IV, the second
objective is reached by  calculating the LV first-order contributions to the
1-loop vacuum polarization. In Sec. V, we give our conclusions and
perspectives.

\section{The theoretical framework\label{sec:2}}

The basic framework of our investigation is a Lorentz-violating (LV) scalar
electrodynamics in which the gauge and scalar fields interact via a
Lorentz-violating nonminimal covariant derivative {including CPT-odd
and CPT-even backgrounds.} The Lagrangian density {describing our
model} is given by
\begin{eqnarray}
\mathcal{L} &=&-\frac{1}{4}F^{\mu \nu }F_{\mu \nu }-\frac{1}{2\xi }(\partial
_{\mu }A^{\mu })^{2}  \label{covnon1} \\[0.2cm]
&&+\left\vert \mathcal{D}_{\mu }\phi \right\vert ^{2}-m^{2}\phi ^{\dag }\phi
-U(|\phi |),  \notag
\end{eqnarray}%
where $F_{\mu \nu }=\partial _{\mu }A_{\nu }-\partial _{\nu }A_{\mu }$, is
the strength-tensor of the electromagnetic field $A_{\mu }$, the term $%
\mathcal{D}_{\mu }\phi $ is the Lorentz-violating nonminimal covariant
derivative,
\begin{equation}
\mathcal{D}_{\mu }\phi =D_{\mu }\phi -\frac{i}{2}g\epsilon _{\mu \nu \alpha
\beta }w^{\nu }{F}^{\alpha \beta }\phi -i\frac{\tilde{g} }{2}{k_{\mu \nu
\alpha \beta }}(\partial^\nu{F^{\alpha \beta}})\phi ,  \label{covnon2}
\end{equation}%
where $D_{\mu }\phi =\partial _{\mu }\phi -ieA_{\mu }\phi $ is the minimal
covariant derivative, {$w^{\nu }$ is a LV vector background providing
CPT-odd terms and $k_{\mu \nu \alpha \beta }$ is a LV tensor background
providing CPT-even terms. The tensor $k_{\mu \nu \alpha \beta }$ possess the
same symmetries as Riemann tensor and null double-trace. The CPT-odd
coupling constant $g$ has mass dimension -1 and the CPT-even one $\tilde{g}$
has mass dimension equal to -2. The explicit first-order LV contributions to
the Lagrangian density (\ref{covnon1}) are given by}
\begin{eqnarray}
\left\vert \mathcal{D}_{\mu }\phi \right\vert ^{2} &=&\left( D_{\mu }\phi
\right) ^{\dag }D^{\mu }\phi  \label{highorder} \\[0.2cm]
&&+\frac{ig}{2}w^{\mu }\epsilon _{\mu \nu \alpha \beta }F^{\alpha \beta }%
\left[ \left( D^{\nu }\phi \right) ^{\dag }\phi -\phi ^{\dag }D^{\nu }\phi %
\right]  \notag \\[0.2cm]
&&-i\frac{\tilde{g} }{2}{k_{\mu \nu \alpha \beta }}{\partial ^{\nu }}{%
F^{\alpha \beta }}\left[ {{{\left( {{D^{\mu }}\phi }\right) }^{\dag }}\phi -{%
\phi ^{\dag }}{D^{\mu }}\phi }\right]  \notag \\[0.2cm]
&&+{\mbox{\it{LV high-orders}}}.  \notag
\end{eqnarray}

{We point out, for our purposes, along the manuscript the charged
scalar field $\phi $ will describe a charged meson, i.e., a pion or a kaon.}
In absence of self-interaction in the scalar sector, the motion equation is
a modified Klein-Gordon one describing charged spinless particle interacting
nonminimally with electromagnetic field. Such interactions produce a shift
on energy-levels of the mesonic atoms, e.g., pionic or kaonic hydrogen. The
second-order LV contributions in Eq. (\ref{highorder}) will not be relevant
for our analysis due to the Coulomb potential is the predominant interaction
and the first-order LV contributions will be treated perturbatively. {%
Besides, due to} the large meson mass, the system can be treated as a
nonrelativistic one via the Breit-Pauli Hamiltonian \cite{thpion} with
additional LV terms.

\section{The nonrelativistic limit and its contribution for the mesonic
hydrogen atoms\label{sec:5}}

The pionic hydrogen is a system where the electron is replaced by a pion. In the relativistic context, it is well established the standard hydrogen atom can be described by Dirac's equation because of the spin-1/2 of the electron. On the other hand, the relativistic treatment of hydrogen pionic can be made in the context of the Klein-Gordon equation due to the pion being spinless. The crucial difference is the fact of the pion is an unstable composite particle constituted by a quark-antiquark pair with mean lifetime around $3.95\times 10^{7}\, \text{eV}^{-1}$. Many properties of the pionic hydrogen were studied in Ref. \cite{thpion} by
considering only QED effects. However, there are quantum chromodynamics (QCD) contributions to the binding energies and level widths of the atomic levels whose most notorious effect is presented by the 1S state. It is measured by means of the X-ray transitions by comparing with the pure electromagnetic bound-state.

We point out some aspects of Lorentz and CPT violation in low-energy QCD has
been analyzed in Ref. \cite{chipt1} and for pions and nucleons in Ref. \cite%
{chipt2}. Both references perform their analysis within the formalism of
chiral perturbation theory \cite{xpt}, an effective quantum field theory
used to describe some low-energy aspects of QCD. Nevertheless, because of
its large mass, the mesonic hydrogen can be considered as a nonrelativistic
system such that we use the transition 2P$\rightarrow $1S or the pure QED
(4P $\rightarrow $3P) ones to impose some upper-bounds for the LV coupling
{constants $g{w}^{\mu }$ and $\tilde{g}k_{\mu \nu \alpha \beta}$}
introduced in Eq. (\ref{covnon2}).

\subsection{The CPT-odd contribution \label{sec:5.1}}

{The relevant term for our analysis is proportional to the CPT-odd
vector background ${w}^{\mu}$ contained in the nonminimal covariant
derivative contribution (\ref{highorder}) which reads}
\begin{equation}
\frac{i}{2}g w_{\mu }\epsilon ^{\mu \nu \alpha \beta} F_{\alpha \beta} \left[
\left( D_{\nu}\phi\right)^{\dag }\phi -\phi ^{\dag}\left(D_{\nu}\phi \right) %
\right],  \label{LscCPToddt}
\end{equation}
whose nonrelativistic limit provides
\begin{equation}
\mathcal{L}_{eff}=g\vec{w}\cdot\vec{B}\phi^{\dag}\phi+...,
\end{equation}
which contributes to the nonrelativistic Hamiltonian with the following
perturbation,
\begin{equation}
\Delta H=-g\vec{w}\cdot\vec{B}.  \label{mmb}
\end{equation}
This interaction is not obtained from standard quantum electrodynamics,
consequently, we will use it to impose some upper-bounds for the coupling $g%
\vec{w}$. The LV coupling $g\vec{w}$ is playing the role of a magnetic
moment for the meson allowing to study it as a background-orbit {%
interaction, then, the correspondent correction to mesonic hydrogen
Hamiltonian can be written as}
\begin{equation}
\Delta H=-\frac{eg\vec{w}\cdot\vec{L}}{4\pi{\mu}r^{3}},  \label{mmb1}
\end{equation}
where $\mu$ is the reduced mass of the proton-meson system and the magnetic
field in Eq. (\ref{mmb}) was set to be
\begin{equation}
\vec{B}=\frac{e\vec{L}}{4\pi{\mu}r^{3}},
\end{equation}
in a similar way occurring in the spin-orbit case.

For convenience, we consider the background $\vec{w}$ along the $z$-axis,
the correspondent shift to the mesonic hydrogen energy is
\begin{equation}
\Delta E=\left\langle \Delta H\right\rangle =-\frac{eg w_{z}}{4\pi\mu }%
\left\langle \frac{L_{z}}{r^{3}}\right\rangle .  \label{mmb2}
\end{equation}
{It indicates only states} with nonnull angular momentum projection $%
L_{z}$ will receive energy corrections, so the energy of the $1S$ state
remains unaltered. All other states gain the following energy correction:
\begin{equation}
\Delta E=-gw_{z}\frac{\mu^{2}e^{7}}{4\pi}\frac{m_{\ell}}{n^{3}\left(
\ell+1\right) \left( \ell+1/2\right) \ell}.  \label{LVshift}
\end{equation}
We observe the states with lower values of $n$ and $\ell $ receive the most
significant Lorentz violating corrections. Consequently, the $2P$ state is
the more affected while $1S$ no receive LV corrections.

\begin{figure}
\centering\includegraphics[width=8cm]{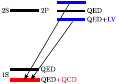}
\caption{The LV affects only states with $l>0$ and split $2P$ in two. The
strong correction shift down the ground state around to $7\,\text{eV}$ for $\protect\pi \mathbf{\text{H}}$ and $\sim 283 \,\text{eV}$ for $K\mathbf{\text{H}}$.}
\label{trans}
\end{figure}

{In the pionic hydrogen case \cite{pionic}, to study of the effects
of the strong interactions are used the measure of the transitions between
excited levels $4P$, $3P$ and $2P$ to the fundamental state $1S$}. The
difference between standard QED predictions and experimental data are
usually considered to compute the shift produced in the $1S$ state by QCD
effects (see Fig. \ref{trans}). The experimental value of the shift is
\begin{equation}
\varepsilon_{1S}\approx(7.086\pm0.007(stat)\pm0.006(sys))\,\text{eV}
\label{epsilonpion}
\end{equation}
to low.

Our purpose is to consider such a energy shift by supposing the LV
corrections are smaller than strong corrections error. Thus, we attain the
following upper-bound,
\begin{equation}
\left\vert {{g_{\pi}}{w_{z}}}\right\vert <1.1\times10^{-12}\,\text{eV}^{-1},
\label{bound1}
\end{equation}
where we use the proton mass to be $938.272081(6)10^{6}\,\text{eV}$ \cite%
{PDG2016}.

Similar procedure can be used with the kaonic hydrogen whose measured strong
shift \cite{kaonic} is,
\begin{equation}
\varepsilon_{1S}\approx -283 \pm 36 (stat) \pm 6 (sys)\,\text{eV},
\label{epsilonkaon}
\end{equation}
such that it provides the upper-bound
\begin{equation}
\left\vert {{g_{K}}{w_{z}}}\right\vert <5.2\times10^{-10}\,\text{eV}^{-1},
\label{bound2}
\end{equation}
which is not better than the one provided by the pionic hydrogen.

A second set of upper-bounds is attained by considering the transition $%
4P\rightarrow 3P$ in pionic hydrogen which corresponds to pure quantum
electrodynamics contribution. {The data in Table 5} of Ref. \cite%
{pionic} allows to estimate the difference between the QED prediction for
transition $4P\rightarrow 3P$ and the experimental measure. Such a
difference is:
\begin{equation}
\left\vert E_{4P\rightarrow 3P}^{{\normalsize {EXP}}}-E_{4P\rightarrow 3P}^{%
{\normalsize {QED}}}\right\vert \approx 0.22\,\text{eV}.  \label{pureele}
\end{equation}%
By putting it in Eq. (\ref{LVshift}), we obtain from the pure QED
contribution the following upper-bound
\begin{equation}
\left\vert {{g_{\pi }}{w_{z}}}\right\vert \leq \mathrm{\ } 3.8\times {%
10^{-11}}\text{eV}^{-1},
\end{equation}
that is close to the previous results obtained when we consider the strong
correction shift.

\subsection{Pure CPT-even contribution \label{sec:5.2}}

{The relevant term for our analysis is the term proportional to the
CPT-even tensor background $k_{\mu \nu \alpha \beta }$ contained in the
nonminimal covariant derivative contribution (\ref{highorder}) which reads}
\begin{equation}
-i\frac{\tilde{g}}{2}{k_{\mu \nu \alpha \beta }}{\partial ^{\nu }}{F^{\alpha
\beta }}\left( {{{\left( {{D^{\mu }}\phi }\right) }^{\dag }}\phi -{\phi
^{\dag }}{D^{\mu }}\phi }\right) ,  \label{LscCPTeven}
\end{equation}%
{whose respective nonrelativistic limit becomes }
\begin{equation}
\mathcal{L}_{eff}=-\tilde{g}{k_{0i0j}}{\partial ^{i}}{E^{j}}{\varphi ^{\ast }%
}\varphi -\frac{\tilde{g}}{2}{k_{0ijk}}{\varepsilon ^{jkl}}{\partial ^{i}}{%
B^{l}}{\varphi ^{\ast }}\varphi +....
\end{equation}%
{By using the decomposition of the tensor $k_{\mu \nu \alpha \beta }$
introduced in Ref. \cite{fermion},}
\begin{equation}
\left( {{\kappa _{DE}}}\right) _{ij}=-2k_{0i0j},\quad {\left( {{\kappa _{DB}}%
}\right) ^{il}}={k^{0ijk}}{\varepsilon ^{jkl}},
\end{equation}%
the contribution to the nonrelativistic Hamiltonian is
\begin{equation}
\Delta H=-\frac{{\tilde{g}}}{2}{\left( {{\kappa _{DE}}}\right) _{ij}}{%
\partial ^{i}}{E^{j}}-\frac{{\tilde{g}}}{2}{\left( {{\kappa _{DB}}}\right)
_{ij}}{\partial ^{i}}{B^{l}}.  \label{mmk}
\end{equation}%
{The second term in (\ref{mmk}) does not contribute at first order
because it is not invariant under parity symmetry:} $\left( \kappa
_{DB}\right) _{ij}\partial ^{i}B^{l}\rightarrow -\left( \kappa _{DB}\right)
_{ij}\partial ^{i}B^{l}$. {On the other hand, the first one do it.
So, by considering the electric field is the one generated by the mesonic
{hydrogen, }}$\left\vert \vec{E}\right\vert \varpropto r^{-2}${%
{, the} contributions to energy of the states $2P$, $3P$ and $4P$,
but not for $1S$ state, are given by}
\begin{equation}
\Delta E_{nP}=-\frac{{\tilde{g}{\alpha ^{7/2}}{\mu ^{3}}\left( \kappa
^{(DE)}\right) }}{{60\sqrt{\pi }{n^{3}}}},  \label{LVshifteven}
\end{equation}%
{where we have introduced the LV quantity $\kappa ^{(DE)}$}
\begin{equation}
\kappa ^{(DE)}=(\kappa _{DE})_{xx}+(\kappa _{DE})_{yy}-2(\kappa _{DE})_{zz}.
\end{equation}

{Despite of the spectral line 2P does not split as it happened in the
CPT-odd case. It is possible to obtain upper-bounds by using the QCD
measurements (\ref{epsilonpion}) and (\ref{epsilonkaon}) for the combination
$\kappa ^{(DE)}$,}
\begin{equation}
\left\vert \tilde{g}_{\pi }\kappa ^{(DE)}\right\vert <1.9\times {10^{-16}}%
\mbox{eV}^{-2},  \label{xxs}
\end{equation}
and
\begin{equation}
\left\vert \tilde{g}_{K}\kappa ^{(DE)}\right\vert <3.3\times {10^{-14}}%
\mbox{eV}^{-2},
\end{equation}
{respectively.}

{By considering the pure electromagnetic interactions $4P\rightarrow
3P$ in pionic hydrogen, it is possible to estimate an upper-bound for $%
\kappa ^{(DE)}$,}
\begin{equation}
\left\vert \tilde{g}_{\pi }\kappa ^{(DE)}\right\vert <1.8\times {10^{-14}} %
\mbox{eV}^{-2},
\end{equation}%
{being two orders of magnitude greater than the previous one (\ref%
{xxs}) obtained by considering only the strong correction.}

\section{1-loop contributions to the photon effective action}

In the remain of manuscript  we will consider the first-order LV contributions at 1-loop effective action of the gauge field of the model (\ref{covnon1}) produced by the CPT-odd and CPT-even backgrounds introduced by the nonminimal covariant derivative (\ref{covnon2}).

The Feynman rules, in the Feynman gauge,  we will use to attain our
goal are depicted below, 

\begin{itemize}
\item Scalar propagator

\begin{figure}[H]
\centering\includegraphics[width=2.5cm]{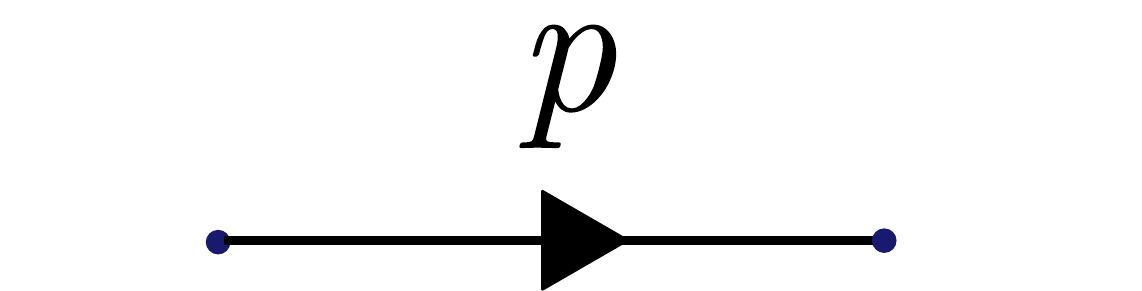}
\end{figure}

\vspace{-0.5cm}

\begin{equation}
i\Delta \left( p \right) = \frac{i}{{{p^2} - {m^2} + i\varepsilon }}.
\end{equation}

\item Photon propagator

\begin{figure}[H]
\centering\includegraphics[width=1.75cm]{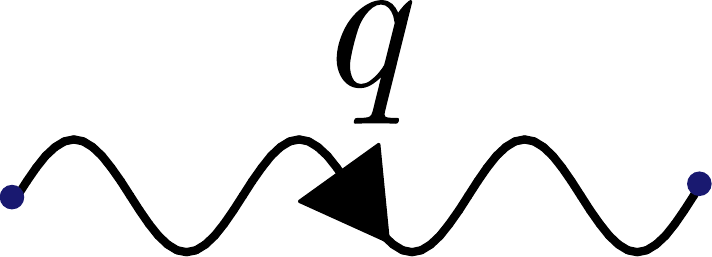}
\end{figure}

\vspace{-0.5cm}

\begin{equation}
i{\Delta _{\mu \nu }}(q) = \frac{- i\eta_{\mu \nu}}{q^2} - \frac{{i\left( {\xi - 1} \right){q_\mu }{q_\nu }}}{{{q^4}}} ,
\end{equation}

\item Tree level scalar-photon 3-vertex

\begin{figure}[H]
\centering\includegraphics[width=4cm]{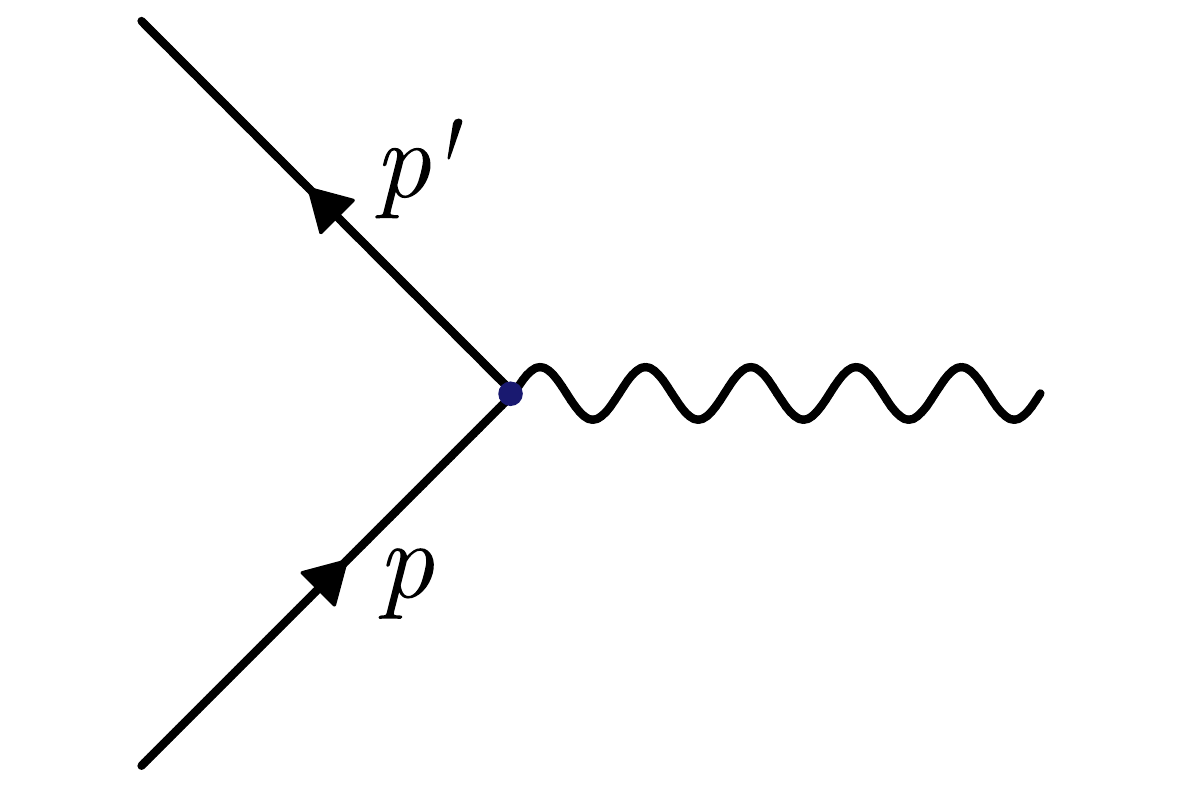}
\end{figure}

\vspace{-0.5cm}

\begin{equation}
i{\Gamma _\mu }\left( {p^{\prime },p} \right) = ie{o_{\mu \alpha }}\left( {%
p^{\prime }- p} \right){\left( {p + p^{\prime }} \right)^\alpha } .
\label{fr3}
\end{equation}

\item Tree level scalar-photon 4-vertex

\begin{figure}[H]
\centering\includegraphics[width=2.8cm]{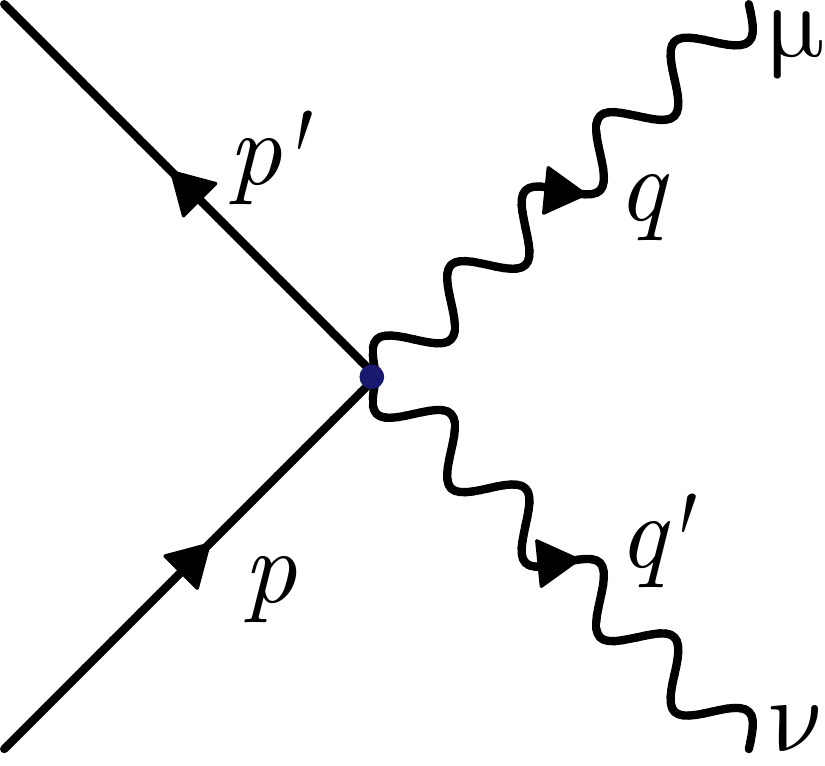}
\end{figure}

\vspace{-0.5cm}

\begin{eqnarray}
i{\Gamma _{\mu \nu }}\left( {p^{\prime },p,q} \right) = i{e^2}{o_{\alpha \mu
}}(q){o^\alpha }_\nu \left( {p - p^{\prime }- q} \right).  \label{fr4}
\end{eqnarray}
\end{itemize}

In Eqs. (\ref{fr3}) and (\ref{fr4}) we have introduced the tensor $o_{\mu
\beta }(q)$ which is given by
\begin{equation}
eo_{\mu \beta }(q) = e{\eta_{\mu \beta }} - ig{\epsilon_{\mu \nu \alpha \beta
}}{w^\nu }{q^\alpha } - \tilde{g} {k_{\mu \nu \alpha \beta }}{q^\nu }{%
q^\alpha }.
\end{equation}

\subsection{1-loop vacuum polarization\label{sec:34}}

{We are interested in the 1-loop radiative corrections containing only first-order contributions of the Lorentz-violating backgrounds $g{w}^{\mu }$ and $\tilde{g}k_{\mu \nu \alpha \beta}$ to the vacuum polarization tensor. The relevant Feynman graphs contributing in the calculus of the
radiative corrections to the vacuum polarization tensor are}

\begin{figure}[H]
\centering\includegraphics[width=6cm]{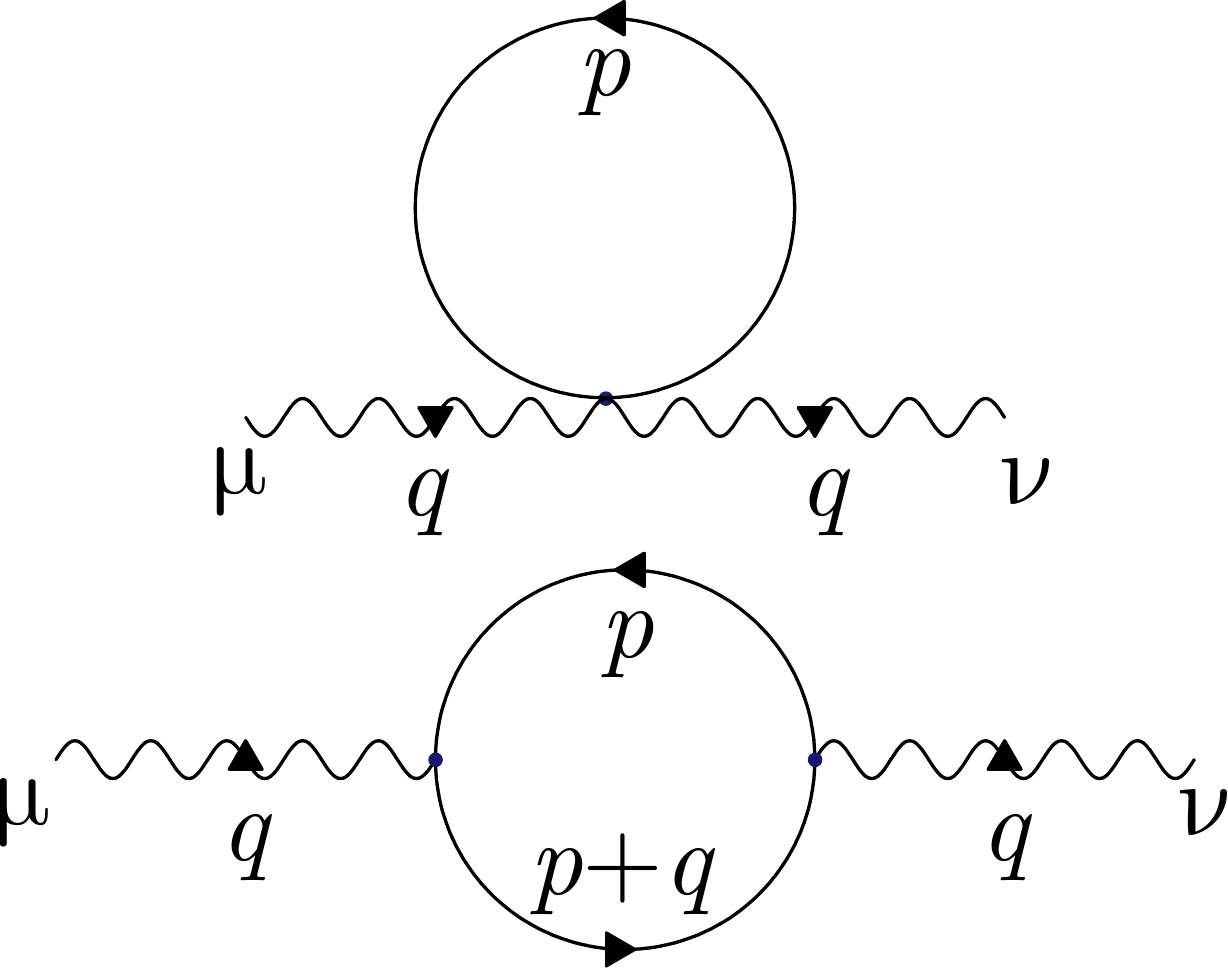}
\caption{1-loop corrections to
propagator.} \label{tvinser}
\end{figure}

By using the Feynman rules, we write the associated amplitude to be
\begin{eqnarray}
-i{\pi _{\mu \nu }}(q) &=&\int {\frac{{{d^{4}}p}}{{{{\left( {2\pi }\right) }%
^{4}}}}i{\Gamma _{\mu \nu }}\left( {p,p,-q}\right) i\Delta \left( p\right) }
\\[0.2cm]
&&\hspace{-1.8cm}+\int {\frac{{{d^{4}}p}}{{{{\left( {2\pi }\right) }^{4}}}}i{%
\Gamma _{\mu }}\left( {p+q,p}\right) i{\Gamma _{\nu }}\left( {p,p+q}\right)
i\Delta \left( {p+q}\right) i\Delta \left( p\right) }.  \notag
\end{eqnarray}%
{Surprisingly, it can be expressed in the following factorized form}
\begin{equation}
-i\pi _{\mu \nu }(q)={o}_{\mu }{}^{\alpha }(q)\left[ -i\pi _{\alpha \beta
}^{sQED}(q)\right] o^{\beta }{}_{\nu }(q), \label{eoo}
\end{equation}%
where $\pi _{\alpha \beta }^{sQED}(q)$ is the 1-loop vacuum polarization
tensor of the usual $sQED$ whose the divergent part (computed via dimensional
regularization technique with $D\rightarrow D-2\epsilon $) is
\begin{equation}
{-i\pi _{\alpha \beta }^{sQED}(q)}=\frac{-ie^{2}}{48\pi ^{2}\epsilon }%
(q^{2}\eta _{\mu \nu }-q_{\mu }q_{\nu })+\mbox{UV finite terms}.
\end{equation}%
{By considering only up to the first-order LV contributions, from (\ref{eoo}) we obtain}
\begin{eqnarray}
\pi _{\mu \nu }(q) &=&\frac{{e^{2}}}{{48{\pi ^{2}}}\epsilon }\left( {{q^{2}}{%
\eta _{\mu \nu }}-{q_{\mu }}{q_{\nu }}}\right)   \label{tp} \\[0.2cm]
&&-i\frac{eg}{{24{\pi ^{2}}}\epsilon }{\epsilon _{\mu \beta \alpha \nu }}{%
w^{\beta }}{q^{\alpha }}{q^{2}}-\frac{e\tilde{g}}{{24{\pi ^{2}}}\epsilon }{%
k_{\mu \beta \alpha \nu }}{q^{\beta }}{q^{\alpha }}{q^{2}}.\quad   \notag
\end{eqnarray}%
{We see the photon 1PI two-point function provides the usual term
plus two Lorentz-violating ones, the first is CPT-odd and the second is
CPT-even. These terms require the original photon action in (\ref{covnon1}) must be modified in order to obtain consistency at 1-loop order to least in the photonic sector.  The new terms can be expressed in the following form,}
\begin{equation}
\Delta \mathcal{L}=\frac{g}{4}{\epsilon_{\mu \beta \alpha \nu }}{u^{\beta}_{(w)}A^{\mu }}\square F^{\alpha \nu }+\tilde{g}{l_{(k)}^{\mu \beta \alpha \nu }}{F_{\mu \beta }}\square {F_{\alpha \nu }}.  \label{lvd56gauge}
\end{equation}%
{These contributions imply in a change in the original structure of photon propagator. A similar term to the CPT-odd one was previously generated in a fermionic model but it was a UV finite radiative correction \cite{Mariz}. Unlike of the nonminimal coupling (\ref{highorder}), the terms in (\ref{lvd56gauge}) belong  to the nonminimal SME. The couplings of the new terms are very limited by experimental data \cite{Data} and for this reason they can be ignored such it was made in the previous section.}

\section{Remarks and conclusions\label{sec:6}}

{We have studied the contributions of the CPT-odd and CPT-even Lorentz-violating nonminimal couplings to the energy spectrum of the mesonic hydrogen and the higher-order radiative corrections engendered by them to the 1-loop photon effective action in a scalar electrodynamics.  The nonminimal LV interactions were introduced by modifying the covariant derivative linking the complex scalar field and the gauge field (see Eq. (\ref{covnon2})).}

{By regarding the nonrelativistic limit, the contributions to the hamiltonian of a mesonic hydrogen atom imply that backgrounds $g\vec{w}$ and  $\tilde{g} k_{\mu\nu\alpha\beta}$ interact directly with the electromagnetic field such as shown in Eqs. (\ref{mmb}) and (\ref{mmk}), respectively. The resulting corrections to the bound-state energy  (see Eqs. (\ref{LVshift}) and (\ref{LVshifteven}), respectively) allow to obtain some upper-bounds for the LV couplings. Such energy-shifts are compared with the experimental data in two cases: The first one, by using the measurement of $1S$ strong-shift correction of the pionic and kaonic hydrogen atoms; and the second one is related to the pure QED transition $4P\rightarrow 3P$ in pionic hydrogen. The obtained results are summarized in the Table \ref{table}.}

\begin{table}[H]
\centering%
\begin{tabular}{ccccc}
\hline\hline
Meson/Atom & \hspace{0.5cm} & 1S shift & \hspace{0.5cm} & $4P\rightarrow 3P$
\\ \hline
$\left\vert {g_{\pi }{w^{\mu }}}\right\vert <$ &  & $1.1\times 10^{-12}\text{%
eV}^{-1}$ &  & $3.8\times {10^{-11}}\text{eV}^{-1}$ \\[0.2cm] \hline
$\left\vert {g_{K}{w^{\mu }}}\right\vert <$ &  & $5.2\times 10^{-10}\text{eV}%
^{-1}$ &  & -- \\[0.2cm] \hline
$\left\vert \tilde{g}_{\pi }\kappa ^{(DE)}\right\vert <$ &  & $1.9\times
10^{-16}$eV$^{-2}$ &  & $1.8\times 10^{-14}$eV$^{-2}$ \\[0.2cm] \hline
$\left\vert \tilde{g}_{K}\kappa ^{(DE)}\right\vert <$ &  & $3.3\times
10^{-14}$eV$^{-2}$ &  & -- \\ \hline\hline
\end{tabular}
\caption{Comparative upper-bounds for the LV couplings.} \label{table}
\end{table}

The radiative corrections at first-order in LV to the 1-loop vacuum polarization modify the structure of the tree-level photon propagator, however, such terms are strongly limited by experimental data \cite{Data}, consequently, they are not considered in our analysis.

Finally, we point out the possibility to study another exotic atoms searching of new effects with the aim to impose more restrictive bounds for the LV parameters. Other possible investigation would be consider the study of the 1-loop renormalization of this model at first-order in Lorentz violation. Advances in these topics will be reported elsewhere.

\begin{acknowledgments}
We thank CAPES, CNPq and FAPEMA (Brazilian agencies) for financial support.
\end{acknowledgments}

\bigskip

\end{document}